\documentclass{IEEEtran}
\usepackage{cite}
\usepackage{amsmath,amssymb,amsfonts}
\usepackage{graphicx}
\usepackage[caption=false,font=footnotesize]{subfig}
\usepackage{textcomp,nicefrac}
\usepackage{xcolor}
\def\BibTeX{{\rm B\kern-.05em{\sc i\kern-.025em b}\kern-.08em
T\kern-.1667em\lower.7ex\hbox{E}\kern-.125emX}}

\begin{document}

\title{Development and Initial Performance of an Upgraded NaI(Tl) Crystal Encapsulation for COSINE-100U}
\author{Doohyeok Lee, Jae Young Cho, Chang Hyon Ha, Eunju Jeon,
Hongjoo Kim, Jinyoung Kim, Kyungwon Kim,
SungHyun Kim, Sun Kee Kim, Won Kyung Kim,
Yeongduk Kim, Young Ju Ko, Hyunseok Lee,
Hyun Su Lee, In Soo Lee, Jaison Lee, Seo Hyun Lee,
Seung Mok Lee, Reina H. Maruyama, Jong-Chul Park,
Kangsoon Park, Kihong Park, Se Dong Park, Kyungmin Seo,
Min Ki Son, and Gyun Ho Yu%
\thanks{Doohyeok Lee, Jae Young Cho, and Hongjoo Kim are with the Department of Physics, Kyungpook National University, Daegu 41566, Republic of Korea.}
\thanks{Chang Hyon Ha and Jinyoung Kim are with the Department of Physics, Chung-Ang University, Seoul 06973, Republic of Korea.}
\thanks{Eunju Jeon, Kyungwon Kim, SungHyun Kim, Won Kyung Kim, Yeongduk Kim, Hyunseok Lee, Hyun Su Lee, In Soo Lee, Jaison Lee, Seo Hyun Lee, Kihong Park, Kyungmin Seo, and Gyun Ho Yu are with the Center for Underground Physics, Institute for Basic Science, Daejeon 34126, Republic of Korea.}
\thanks{Won Kyung Kim, Yeongduk Kim, Hyunseok Lee, Hyun Su Lee, and Seo Hyun Lee are also with the IBS School, University of Science and Technology, Daejeon 34113, Republic of Korea.}
\thanks{Sun Kee Kim is with the Department of Physics, Seoul National University, Seoul 08826, Republic of Korea.}
\thanks{Young Ju Ko is with the Department of Physics, Jeju National University, Jeju 63243, Republic of Korea.}
\thanks{Seung Mok Lee is with the Department of Physics, Mellon College of Science, Carnegie Mellon University, Pittsburgh, PA 15213, USA.}
\thanks{Reina H. Maruyama is with the Department of Physics and Wright Laboratory, Yale University, New Haven, CT 06520, USA.}
\thanks{Jong-Chul Park and Min Ki Son are with the Department of Physics and IQS, Chungnam National University, Daejeon 34134, Republic of Korea.}
\thanks{Kangsoon Park is with the Center for Yemilab Operation, Institute for Basic Science, Jeongseon 26141, Republic of Korea.}
\thanks{Corresponding author: In Soo Lee (e-mail: islee@ibs.re.kr).}
}

\maketitle

\begin{abstract}
The COSINE-100 experiment was designed to test the DAMA/LIBRA annual-modulation claim using low-background NaI(Tl) detectors. For the COSINE-100U upgrade, we developed a new crystal-encapsulation system to increase light-collection efficiency while preserving long-term detector stability, thereby improving sensitivity to low-mass dark matter. The upgraded design eliminates the quartz optical windows used in COSINE-100 and directly couples the photomultiplier tubes (PMTs) to the crystal end faces through 2-mm-thick silicone optical pads, thereby reducing the number of optical interfaces. For the larger crystals, the crystal edges were beveled to guide scintillation light more efficiently onto 3-inch high-quantum-efficiency PMTs. The performance study uses 2462~h (102.6~days) of room-temperature COSINE-100U data and, for direct background comparisons, reference COSINE-100 data acquired near the end of operation. 698~h (29.1~days) of COSINE-100 data acquired near the end of operation in March 2023. All eight crystals showed higher light yields than in COSINE-100, with values ranging from 15.8 to 27.7~p.e./keV; six crystals exceeded 20~p.e./keV. The measured bulk-$\alpha$ rates were lower than the COSINE-100 values and consistent with the expected time evolution of internal $^{210}$Pb, while the 1--2-MeV surface-$\alpha$ rates were substantially reduced. The upgrade also restored two crystals that had previously been excluded from the COSINE-100 physics analysis because of poor optical performance. Independent validation tests demonstrated that the encapsulation remains mechanically robust and optically stable during long-term immersion in liquid scintillator at low temperature. This paper presents the encapsulation design, the room-temperature detector performance, and the reduction in surface-related backgrounds achieved at the Yemilab facility.
\end{abstract}
\begin{IEEEkeywords}
Dark matter, NaI(Tl), scintillation detector, crystal encapsulation, low-background experiment
\end{IEEEkeywords}

\section{Introduction}
\label{sec:introduction}

\IEEEPARstart{E}{xtensive} astrophysical and cosmological
observations indicate that dark matter constitutes a substantial
fraction of the matter in the Universe, yet its particle nature
and interaction mechanisms remain unknown~\cite{
Clowe:2006eq,Planck:2018vyg,Bertone:2016nfn}.
Despite extensive direct-detection searches, no dark-matter
signal has been independently confirmed~\cite{
MarrodanUndagoitia:2015veg,Schumann:2019eaa}.
The DAMA/LIBRA experiment, which operates an approximately
250-kg array of NaI(Tl) scintillators, has reported a persistent
annual modulation that it interprets as evidence for dark-matter
interactions~\cite{
Bernabei:1998fta,Bernabei:2013xsa,Bernabei:2018yyw,
Bernabei:2021kdo}.
Because the origin of this signal remains unresolved, an
independent test using the same NaI(Tl) target material is
essential for distinguishing a possible dark-matter signal from
detector- or material-specific effects~\cite{
Schumann:2019eaa,ParticleDataGroup:2022pth}.

The COSINE-100 experiment was designed to perform such a test.
It operated a 106-kg array of NaI(Tl) detectors at the Yangyang
Underground Laboratory (Y2L) for more than six years~\cite{
Adhikari:2017esn}.
Spectral analyses of the COSINE-100 data found no evidence
supporting standard weakly interacting massive particle (WIMP)
interpretations of the DAMA/LIBRA signal~\cite{
Adhikari:2018ljm,COSINE-100:2021xqn,COSINE-100:2025xqn}.
Using the full data set, COSINE-100 also measured an
annual-modulation amplitude that was inconsistent with the
DAMA/LIBRA result at more than $3\sigma$~\cite{
COSINE-100:2024jkg}.
Its sensitivity at the lowest energies, however, was limited by
the optical performance of the detectors. The crystals in the
original configuration produced an average light yield of
approximately 15 photoelectrons (PE) per keV, which constrained
the achievable analysis threshold and reduced sensitivity to
low-mass dark-matter candidates~\cite{
Adhikari:2017esn,lee2025upgrading}.
Improving light collection and reducing low-energy backgrounds
were therefore central goals of the next experimental phase.

After COSINE-100 completed data taking in March 2023, the
detector was upgraded to COSINE-100U and relocated to Yemilab,
a newly constructed deep-underground laboratory in
Korea~\cite{lee2025upgrading,Park:2024sio}.
Stable room-temperature physics data taking began in September
2025, followed by a dedicated calibration campaign. Physics data
taking at the nominal operating temperature of $-30\,^{\circ}\mathrm{C}$
began in May 2026.

In this paper, we present the initial performance of the complete
COSINE-100U detector array, which comprises eight NaI(Tl)
crystals with a total mass of 99.1~kg~\cite{lee2025upgrading}.
The analysis is based on 2462~h (102.6~days) of stable
room-temperature physics data acquired at Yemilab and focuses on
the measured light yields, bulk- and surface-$\alpha$ rates, and
single-hit energy spectra. Where applicable, the results are compared with 698~h (29.1~days) of COSINE-100 data acquired near the end of operation in March 2023. The performance of the
detector at $-30\,^{\circ}\mathrm{C}$ is beyond the scope of the
present study.

\section{Upgraded Encapsulation Design and Crystal Machining}
\label{sec:encapsulation}
\begin{figure}[!t]
    \centering
    \includegraphics[width=\columnwidth]{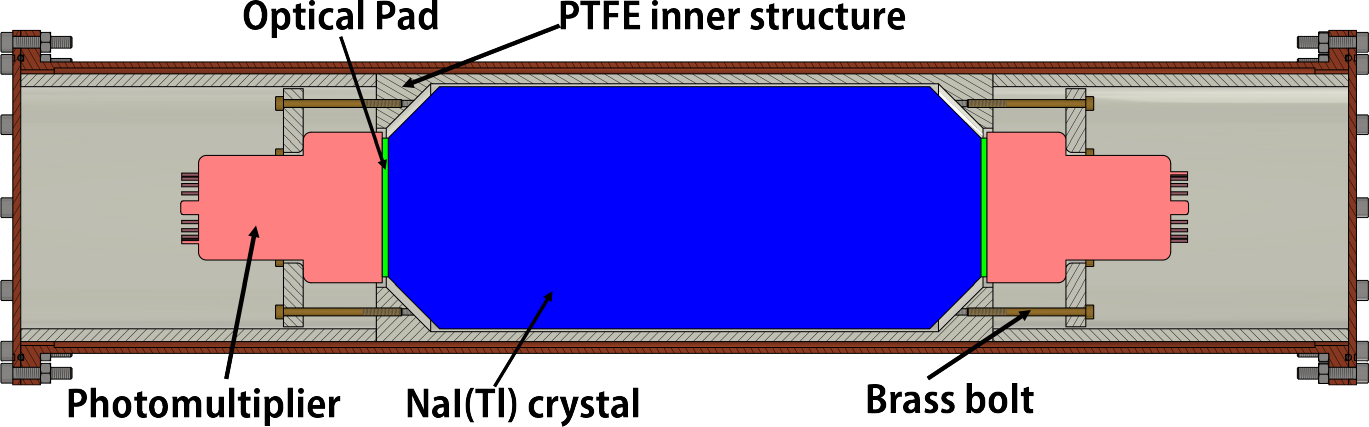}
    \caption{Encapsulation design of the COSINE-100U C6 crystal. 
The 4.8-inch-diameter NaI(Tl) crystal has $45^{\circ}$ beveled
edges that guide scintillation light toward the 3-inch PMTs.
The crystal is wrapped with a PTFE reflector, and each PMT is
coupled directly to a crystal end face through a 2-mm-thick
silicone optical pad. A rigid 5-mm-thick PTFE inner structure
supports the crystal--PMT assembly, which is hermetically
enclosed in an OFC housing to prevent contact with the
surrounding liquid scintillator.}
    \label{fig:encapsulation}
\end{figure}

The upgrade required a substantial redesign of the crystal encapsulation to improve photon transport while preserving hermeticity~\cite{Choi:2020qcj,NEON:2022hbk,Choi:2024trx}. In the original COSINE-100 design, each crystal was optically coupled to the PMTs through 12-mm-thick quartz windows, which introduced multiple optical interfaces and associated photon losses~\cite{Adhikari:2017esn,Choi:2020qcj}. Figure~\ref{fig:encapsulation} shows a schematic of the COSINE-100U encapsulation design. The quartz windows were removed, and the PMTs were coupled directly to the crystal end faces through 2-mm-thick silicone optical pads. For crystals whose diameters exceeded the 3-inch PMT photocathode diameter, the crystal edges were machined with $45^{\circ}$ bevels to redirect scintillation photons toward the PMTs. The crystal--PMT assembly is supported by a rigid polytetrafluoroethylene (PTFE) inner structure and hermetically sealed inside an oxygen-free copper (OFC) housing to prevent contact with the surrounding liquid scintillator~\cite{Choi:2020qcj,lee2025upgrading,Choi:2024trx}.

Because NaI(Tl) is highly hygroscopic, crystal machining, polishing, and assembly were performed under controlled low-humidity conditions. After the bevels were machined, the exposed crystal surfaces were repolished and cleaned before the crystals were wrapped with PTFE reflector sheets. Each crystal was then mounted together with two PMTs in the rigid PTFE inner support and enclosed in the OFC housing. These controlled handling and surface-treatment procedures were intended to
prevent moisture exposure and minimize the introduction of surface contamination during re-encapsulation.

A similar direct-coupling configuration implemented in the NEON experiment increased the light yield by approximately 50\% and maintained stable detector performance for more than two years~\cite{Choi:2024trx}. Independent measurements have also shown that cooling NaI(Tl)
crystals to approximately $-35\,^{\circ}\mathrm{C}$ increases their intrinsic light yield and may improve pulse-shape discrimination~\cite{Lee:2021aoi}. Accordingly, the COSINE-100U encapsulation was validated through prolonged immersion in liquid scintillator at low temperature before deployment in the low-temperature COSINE-100U configuration~\cite{Park_2026}.

\section{Experimental Setup}
\label{sec:setup}

\subsection{Yemilab Facility}
\label{subsec:yemilab}

The COSINE-100U experiment is located at Yemilab in Jeongseon, Gangwon Province, Republic of Korea. The laboratory is situated approximately 1,000~m underground~\cite{Park:2024sio}. The measured cosmic-muon flux is approximately $1.0\times10^{-7}$~muons~cm$^{-2}$~s$^{-1}$, corresponding to approximately one quarter of the flux measured at Y2L~\cite{Park:2024sio}. The cavern environment is also controlled to limit radioactive and particulate contamination. The ambient radon concentration is maintained below 50~Bq~m$^{-3}$ by a high-flow ventilation system, while the Radon Reduction System can supply air with a radon concentration below 100~mBq~m$^{-3}$. Epoxy-coated floors and controlled access help maintain the PM10 concentration below 10~$\mu$g~m$^{-3}$~\cite{Park:2024sio}.

To house the COSINE-100U detector setup, a dedicated walk-in low-temperature room equipped with a 10-kW cooling system was constructed inside the experimental cavern. The room is designed to maintain the complete detector and liquid-scintillator veto system at a stable operating temperature of $-30\,^{\circ}\mathrm{C}$~\cite{Park:2024sio,lee2025upgrading}. Cooling is expected to increase the intrinsic scintillation light yield of the NaI(Tl) crystals and may improve pulse-shape discrimination for low-energy nuclear recoils. These improvements are expected to extend the sensitivity of COSINE-100U to low-mass dark matter~\cite{Lee:2021aoi,lee2025upgrading}.

\begin{figure}[!t]
    \centering
    \subfloat[]{\includegraphics[width=0.95\linewidth]{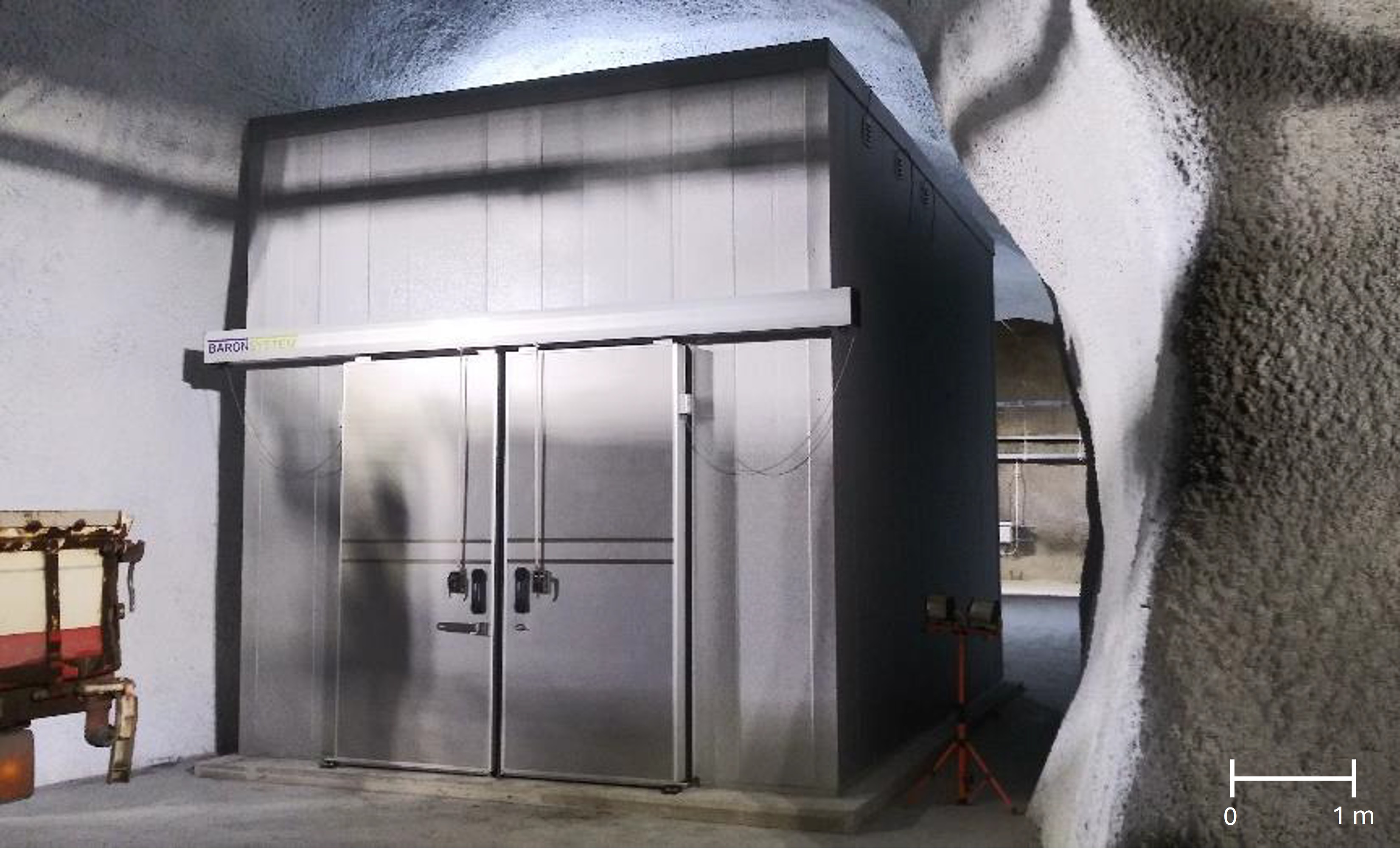}\label{fig:fridgeroom}}\\
    \subfloat[]{\includegraphics[width=0.95\linewidth]{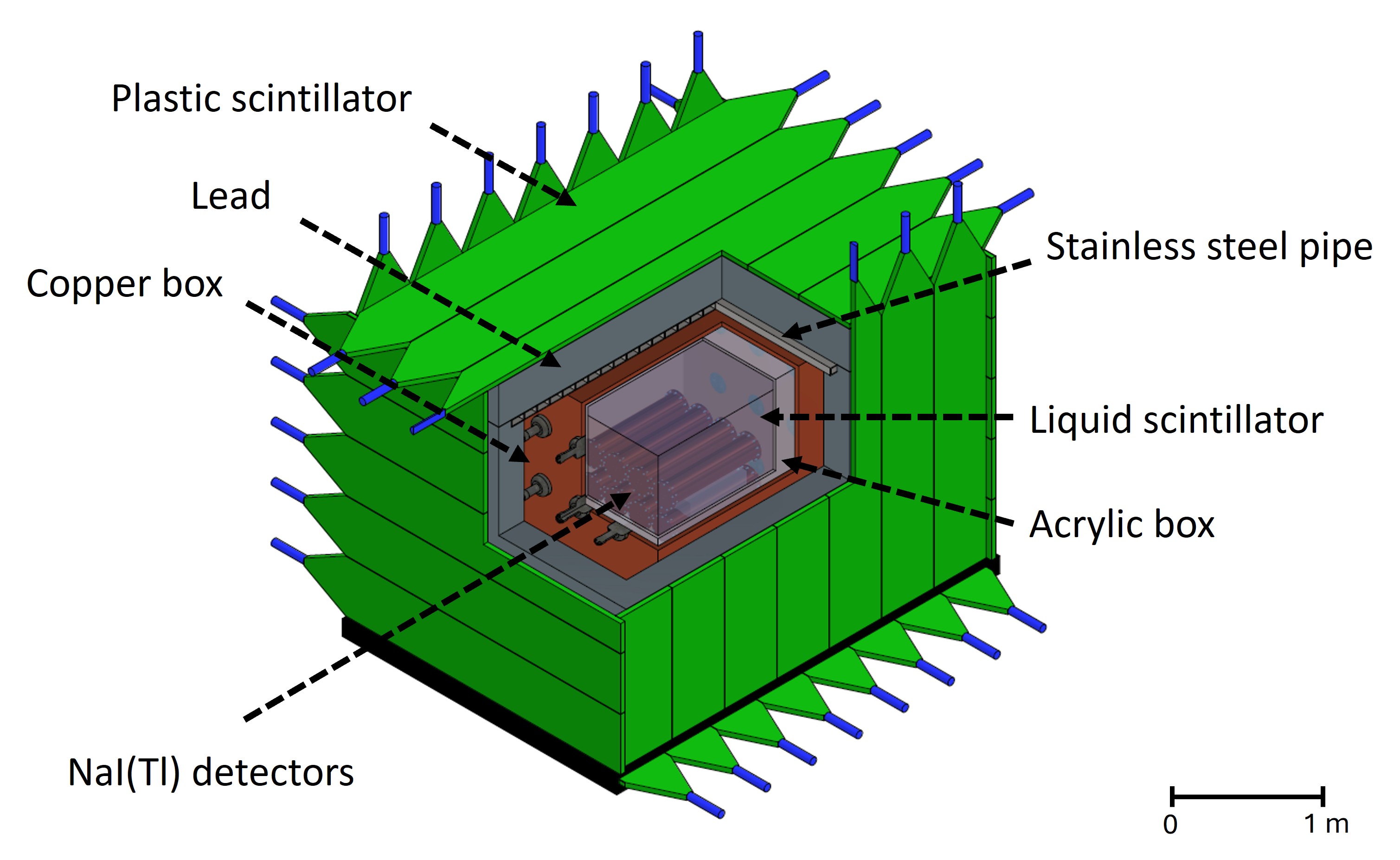}\label{fig:shield}}
    \caption{COSINE-100U at Yemilab. (a) Dedicated low-temperature room designed for operation at $-30\,^{\circ}\mathrm{C}$. (b) Schematic of the detector and multilayer shielding. The setup contains eight NaI(Tl) crystals with a total mass of 99.1~kg immersed in 2,200~L of liquid scintillator and surrounded by 3~cm of copper, 20~cm of lead, and 3~cm of plastic scintillator.}
    \label{fig:yemilab}
\end{figure}

\subsection{Shielding Design}
\label{subsec:shielding}

The COSINE-100U shielding system is installed inside the low-temperature room and reuses most of the shielding components from COSINE-100~\cite{lee2025upgrading}. From the innermost layer outward, the detector is surrounded by 2,200~L of liquid scintillator, 3~cm of copper, 20~cm of lead, and 3~cm of plastic scintillator. The liquid- and plastic-scintillator systems serve as active vetoes that tag events associated with radioactive backgrounds and cosmic-ray muons~\cite{lee2025upgrading}.

The original COSINE-100 shield used an internal steel support structure of approximately 4~t to support the lead and provide a mechanical opening system. This structure was removed in COSINE-100U to reduce the amount of material near the detector~\cite{lee2025upgrading}. Instead, following an approach similar to that used in NEON, the lead bricks are stacked on a precisely leveled steel base plate, and the upper shielding is supported by 180-cm-long stainless-steel rectangular tubes, each with a cross section of 5~cm $\times$ 10~cm~\cite{lee2025upgrading}. Figures~\ref{fig:fridgeroom} and~\ref{fig:shield} show the low-temperature room and the COSINE-100U shielding configuration, respectively.

\subsection{Data Acquisition System}
\label{subsec:daq}

The COSINE-100U data acquisition (DAQ) system follows the architecture used successfully in COSINE-100 and NEON~\cite{Adhikari:2017esn,NEON:2022hbk}. It records signals from 16 PMTs coupled to the eight NaI(Tl) crystals and 18 PMTs that read out the liquid scintillator (LS). Each crystal PMT provides two readout channels: a high-gain anode channel for low-energy events and a low-gain dynode channel for high-energy events. The anode and dynode signals are amplified by factors of 30 and 100, respectively, using custom preamplifiers. The amplified crystal signals are digitized by 12-bit flash analog-to-digital converters (FADCs) operating at 500~MS/s, whereas the unamplified LS signals are digitized by 12-bit ADCs (SADCs) operating at 62.5~MS/s.

A crystal trigger is issued when both anode channels of the same crystal exceed a threshold corresponding to approximately one photoelectron within a 200-ns coincidence window. Channel-level triggers are generated by field-programmable gate arrays in the FADCs, and a central trigger and clock board synchronizes the final trigger decision across all modules. When any crystal satisfies the trigger condition, all FADC and SADC modules record the event. Each FADC channel records an 8-$\mu$s-long waveform, including a 2.4-$\mu$s pretrigger interval. The event data are transferred to the DAQ computer through a USB~3.0 interface and written to ROOT files for offline analysis.

\subsection{Detector Operation}
\label{subsec:operation}

Stable COSINE-100U physics data taking began at Yemilab in September 2025 at a room temperature of approximately $23\,^{\circ}\mathrm{C}$. The room-temperature run was followed by a two-month calibration campaign using a $^{22}$Na source to characterize the energy scale and detector response of the upgraded crystals. After completion of the calibration campaign, the detector was prepared for low-temperature operation. Physics data taking at the nominal operating temperature of $-30\,^{\circ}\mathrm{C}$ began in May 2026. The room-temperature data set used in this study corresponds to 2462~h (102.6~days) of live time. For comparisons with the previous detector configuration, we use 6998~h (29.1~days) of COSINE-100 data acquired near the end of operation in March 2023.

\section{Data Analysis}
\label{sec:analysis}

\subsection{Light-Yield Measurement}
\label{subsec:light_yield}

\begin{figure*}[!t]
    \centering
    \includegraphics[width=0.47\textwidth]{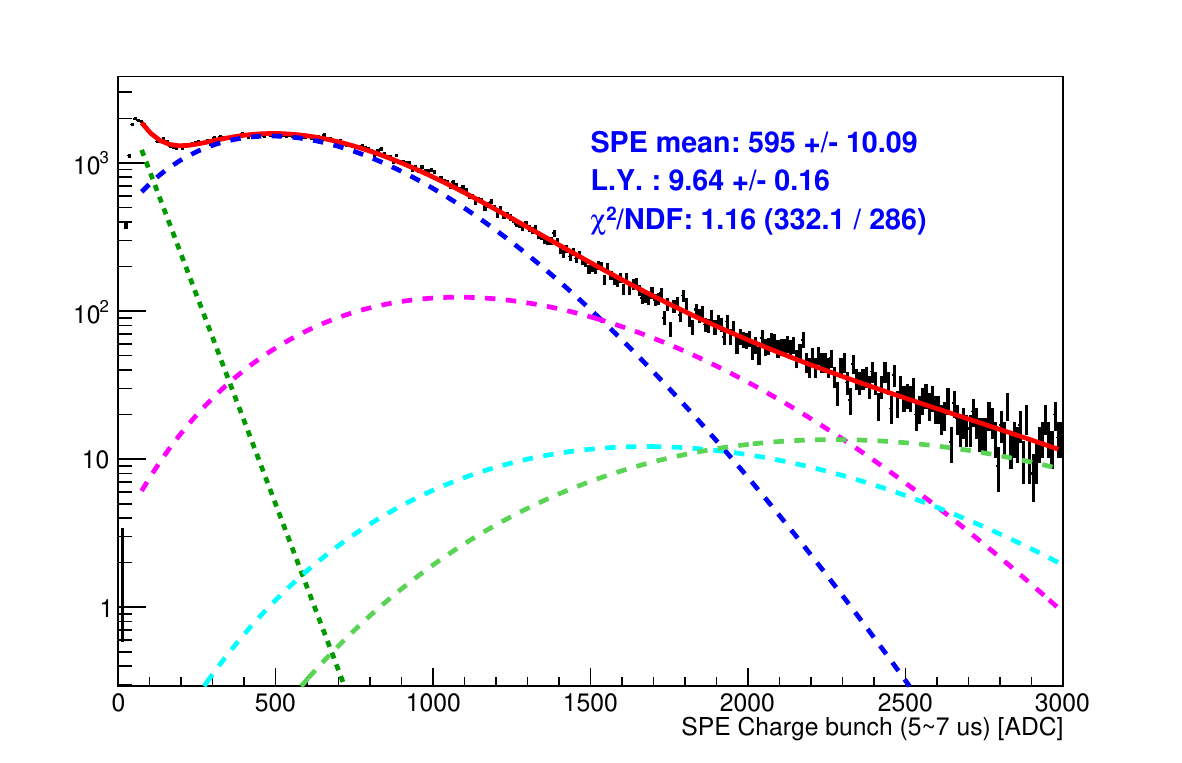}
    \includegraphics[width=0.47\textwidth]{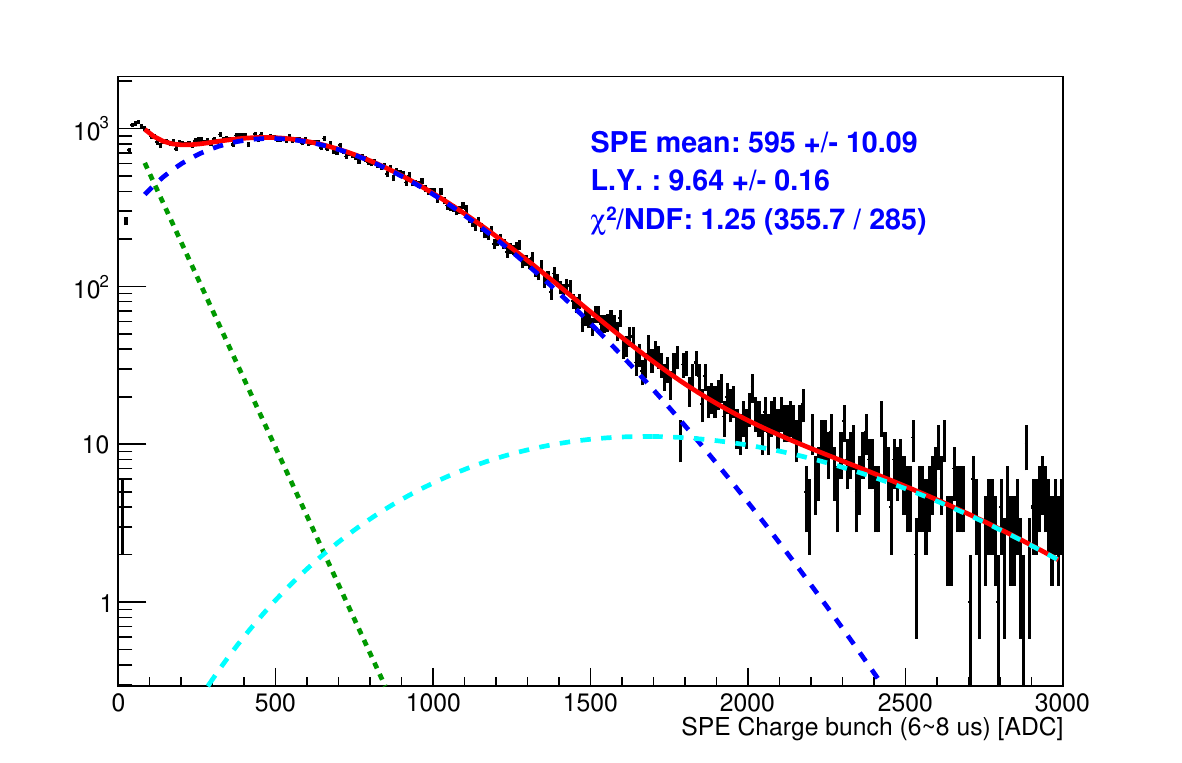}
    \caption{
    \textbf{Determination of the mean single-photoelectron (SPE) charge for Crystal~6.} The charge distributions of isolated clusters in the 5--7~$\mu$s (left) and 6--8~$\mu$s (right) windows are fitted simultaneously with a model including one-, two-, three-, and four-photoelectron components. The fit yields a mean SPE charge of 595.6~ADC units. The error bars indicate the standard deviation of the values contributing to each plotted point.
    }
    \label{fig:SPE}
\end{figure*}

The light-collection efficiency of each COSINE-100U crystal was measured using the 59.54-keV gamma-ray line from a $^{241}$Am source. The procedure follows the approach used in previous NaI(Tl) detector studies~\cite{Choi:2020qcj,Choi:2024ziz,lee2025upgrading}. First, the mean single-photoelectron (SPE) charge was determined from isolated photoelectron clusters in the tail of the 59.54-keV scintillation pulse, where contamination from overlapping photoelectrons is reduced. Figure~\ref{fig:SPE} shows the SPE-charge fits for Crystal~6. The charge distributions in the 5--7~$\mu$s and 6--8~$\mu$s windows were fitted simultaneously with a model including one-, two-, three-, and four-photoelectron components. The integrated charge of the 59.54-keV peak was then divided by the mean SPE charge to determine the number of photoelectrons (NPE). The light yield was calculated by dividing the NPE by 59.54~keV.


Table~\ref{tab:light_yield} compares the crystal masses and measured light yields before and after the upgrade. All eight crystals showed substantial improvements. Crystal~3 achieved the highest light yield, increasing from $15.5\pm1.6$ to $27.7\pm0.5$~p.e./keV. Crystals~5 and 8, whose optical performance had previously degraded to $7.3\pm0.7$ and $3.5\pm0.3$~p.e./keV, recovered to $17.8\pm0.5$ and $15.8\pm0.5$~p.e./keV, respectively. Six of the eight crystals achieved light yields above 20~p.e./keV. These results demonstrate that the upgraded optical coupling substantially improved the optical performance of the array. The reduced crystal masses in Table~\ref{tab:light_yield} result from surface polishing and bevel machining during re-encapsulation.


\begin{table}[!t]
\centering
\caption{Comparison of crystal masses and measured light yields between the original COSINE-100 and upgraded COSINE-100U configurations.}
\label{tab:light_yield}
\renewcommand{\arraystretch}{1.3}
\resizebox{\columnwidth}{!}{%
\begin{tabular}{c|cc|cc}
\hline\hline
\textbf{Crystal} & \multicolumn{2}{c|}{\textbf{Mass (kg)}} & \multicolumn{2}{c}{\textbf{Light yield (p.e./keV)}} \\ \cline{2-5}
\textbf{\#} & COSINE-100 & COSINE-100U & COSINE-100 & COSINE-100U \\ \hline
1 & 8.3  & 7.1  & $14.9 \pm 1.5$ & $24.5 \pm 0.9$ \\
2 & 9.2  & 8.7  & $14.6 \pm 1.5$ & $24.9 \pm 0.5$ \\
3 & 9.2  & 8.7  & $15.5 \pm 1.6$ & $27.7 \pm 0.5$ \\
4 & 18.0 & 16.9 & $14.9 \pm 1.5$ & $22.6 \pm 0.5$ \\
5 & 18.3 & 17.2 & $7.3 \pm 0.7$  & $17.8 \pm 0.5$ \\
6 & 12.5 & 11.6 & $14.6 \pm 1.5$ & $20.9 \pm 0.6$ \\
7 & 12.5 & 11.6 & $14.0 \pm 1.4$ & $22.5 \pm 0.6$ \\
8 & 18.3 & 17.2 & $3.5 \pm 0.3$  & $15.8 \pm 0.5$ \\
\hline\hline
\end{tabular}%
}
\end{table}

\subsection{Internal Alpha-Activity Measurements}
\label{subsec:alpha}

\begin{figure}[!t]
    \centering
    \includegraphics[width=\linewidth]{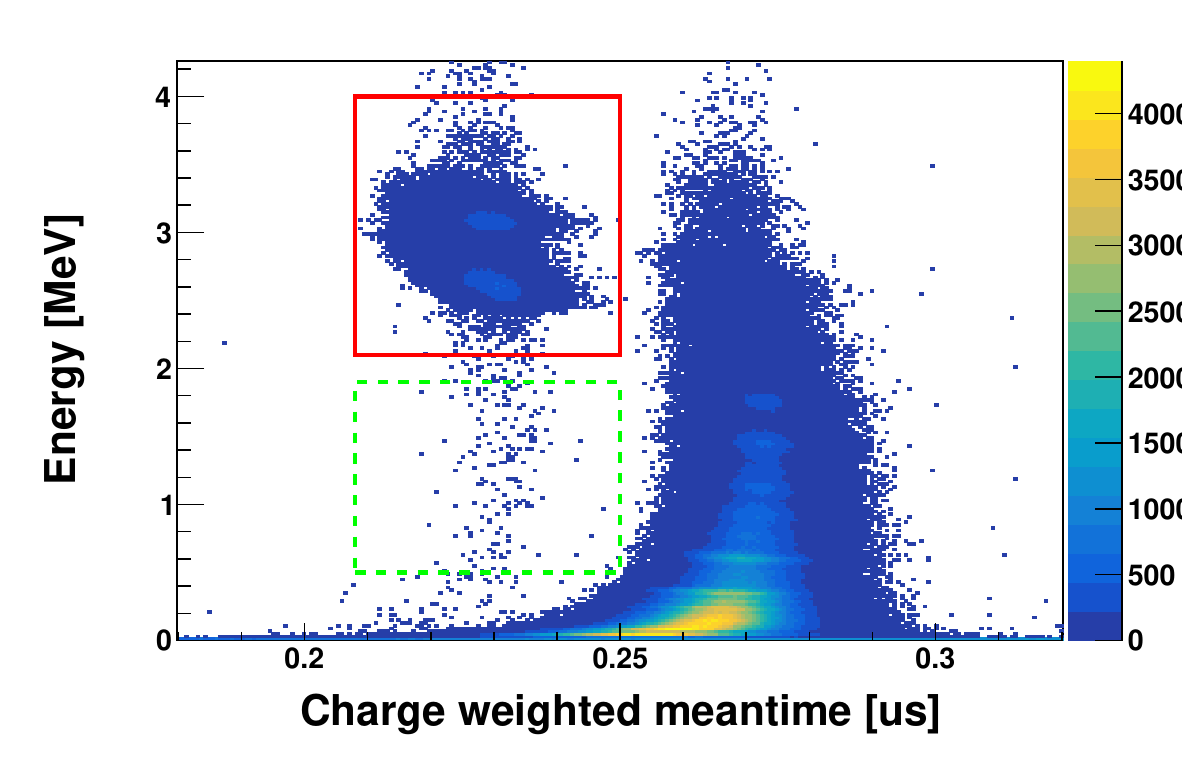}
    \caption{\textbf{Selection of $\alpha$ events in Crystal~6.} The charge-weighted mean decay time is shown as a function of measured energy. The $\alpha$ population has shorter decay times than the $\beta/\gamma$ band. The red solid box indicates bulk-$\alpha$ events dominated by internal $^{210}$Po decays, and the green dashed box indicates surface-$\alpha$ events in the 1--2-MeV energy range.}
    \label{fig:alpha}
\end{figure}

To measure the internal $\alpha$ activity, we used the charge-weighted mean decay time to distinguish $\alpha$-induced events from $\beta/\gamma$ events,
\begin{equation}
\langle t \rangle = \frac{\sum_i A_i t_i}{\sum_i A_i},
\end{equation}
where $A_i$ and $t_i$ are the charge and time of the $i$th digitized waveform bin, respectively. As shown in Fig.~\ref{fig:alpha}, the $\alpha$ events form a distinct population at shorter mean decay times than the $\beta/\gamma$ events. The red solid box identifies the bulk-$\alpha$ population, which is dominated by internal $^{210}$Po decays, whereas the green dashed box identifies the low-energy surface-$\alpha$ population in the measured energy range of 1--2~MeV.

The bulk-$\alpha$ population originates primarily from $^{210}$Po decays supported by internal $^{210}$Pb contamination introduced during crystal growth. Because $^{210}$Pb has a half-life of 22.3~years, the corresponding activity is expected to decrease slowly with time. The bulk-$\alpha$ rates in Table~\ref{tab:alpha_rate} were obtained from 698~h (29.1~days) of COSINE-100 data acquired near shutdown in March 2023 and 2462~h (102.6~days) of room-temperature COSINE-100U data, with the rates normalized by crystal mass. For all six crystals with direct measurements in both configurations, the COSINE-100U bulk-$\alpha$ rates are lower than the COSINE-100 values. The observed decrease is consistent with the expected time evolution of the internal $^{210}$Pb activity and indicates that the re-encapsulation process did not introduce detectable additional bulk contamination.

Surface-$\alpha$ events can arise from contamination on the crystal surfaces or on the PTFE reflector during crystal handling and encapsulation. We use the event rate in the 1--2-MeV region as an indicator of surface contamination and normalize the rate to the exposed crystal surface area. During re-encapsulation, the crystals were handled in a humidity- and radon-controlled glove box, polished, and cleaned with anhydrous ethanol and isopropanol. As summarized in Table~\ref{tab:alpha_rate}, the surface-$\alpha$ rates are lower in COSINE-100U for all six crystals with a direct COSINE-100 comparison, with reductions approaching an order of magnitude for Crystals~3, 6, and 7. Crystals~5 and 8, which were not used in the previous COSINE-100 physics analysis because of poor optical performance, also exhibit low surface-$\alpha$ rates after re-encapsulation. These results show that the controlled encapsulation and surface-treatment procedures effectively limited surface contamination. Further controlled studies using sample crystals and varied surface-treatment procedures are planned to investigate the origin of the remaining surface-$\alpha$ activity.

\begin{table}[!t]
\centering
\caption{Comparison of bulk- and surface-$\alpha$ rates. The COSINE-100 values are based on 698~h (29.1~days) of data acquired near shutdown in March 2023, and the COSINE-100U values are based on 2462~h (102.6~days) of room-temperature data. Bulk rates are normalized by crystal mass, and surface rates are normalized by exposed crystal area. The quoted uncertainties are statistical only.}
\label{tab:alpha_rate}
\renewcommand{\arraystretch}{1.3}
\resizebox{\columnwidth}{!}{%
\begin{tabular}{c|cc|cc}
\hline\hline
\textbf{Crystal} & \multicolumn{2}{c|}{\textbf{Bulk $\alpha$ (mBq/kg)}} & \multicolumn{2}{c}{\textbf{Surface $\alpha$ (nBq/cm$^2$)}} \\ \cline{2-5}
\textbf{\#} & COSINE-100 & COSINE-100U & COSINE-100 & COSINE-100U \\ \hline
1 & $2.59 \pm 0.01$ & $2.39 \pm 0.01$ & $48.78 \pm 4.49$ & $30.72 \pm 1.83$ \\
2 & $1.69 \pm 0.01$ & $1.56 \pm 0.01$ & $50.32 \pm 4.24$ & $21.74 \pm 1.46$ \\
3 & $0.63 \pm 0.01$ & $0.55 \pm 0.01$ & $102.42 \pm 6.05$ & $16.77 \pm 1.28$ \\
4 & $0.64 \pm 0.01$ & $0.58 \pm 0.01$ & $34.42 \pm 2.76$ & $12.98 \pm 0.88$ \\
5 & -- & $1.56 \pm 0.01$ & -- & $11.06 \pm 0.81$ \\
6 & $1.53 \pm 0.01$ & $1.36 \pm 0.01$ & $133.09 \pm 6.19$ & $15.06 \pm 1.08$ \\
7 & $1.51 \pm 0.01$ & $1.37 \pm 0.01$ & $125.60 \pm 6.02$ & $17.79 \pm 1.18$ \\
8 & -- & $1.57 \pm 0.01$ & -- & $13.56 \pm 0.90$ \\
\hline\hline
\end{tabular}%
}
\end{table}

\subsection{Single-Hit Energy Spectrum}
\label{subsec:single_hit}

 A single-hit event is defined as an event with energy deposition in only one NaI(Tl) crystal, no coincident hit in any other crystal, and no coincident LS energy deposition above 80~keV. The same 80-keV LS veto threshold was applied to both COSINE-100 and COSINE-100U.


PMT-induced noise and alpha-like pulses were rejected using several pulse-shape-discrimination variables. The primary timing variable was the PMT-based mean-time parameter,
\begin{equation}
\mathrm{pmpar}=\log_{10}\!\left[
\left(\frac{\langle t\rangle_{0}}{1~\mu\mathrm{s}}\right)
\left(\frac{\langle t\rangle_{1}}{1~\mu\mathrm{s}}\right)
\right],
\end{equation}
where $\langle t\rangle_{0}$ and $\langle t\rangle_{1}$ are the charge-weighted mean times measured by the two PMTs coupled to a crystal within a 500-ns window. Combining the two PMTs in a single parameter suppresses events in which either PMT contains an anomalously fast pulse. Candidate scintillation events were required to satisfy $-1.9<\mathrm{pmpar}<-1.0$, corresponding to geometric-mean times of approximately 0.11--0.32~$\mu$s under the present operating conditions.

Two additional variables were constructed from charge ratios. The variable $x_1$ is the fraction of charge in the 100--600-ns interval relative to the total charge in the 0--600-ns window, whereas $x_2$ is the fraction in the prompt 0--50-ns interval relative to the same total. The requirements $0.50<x_1<0.95$ and $0<x_2<0.30$ retain the slower NaI(Tl) scintillation pulses while rejecting fast noise concentrated near the beginning of the waveform. A crystal-dependent two-PMT charge-asymmetry requirement was also applied to suppress position-dependent and PMT-related backgrounds.

\begin{figure*}[!t]
   \centering
   \includegraphics[width=0.23\textwidth]{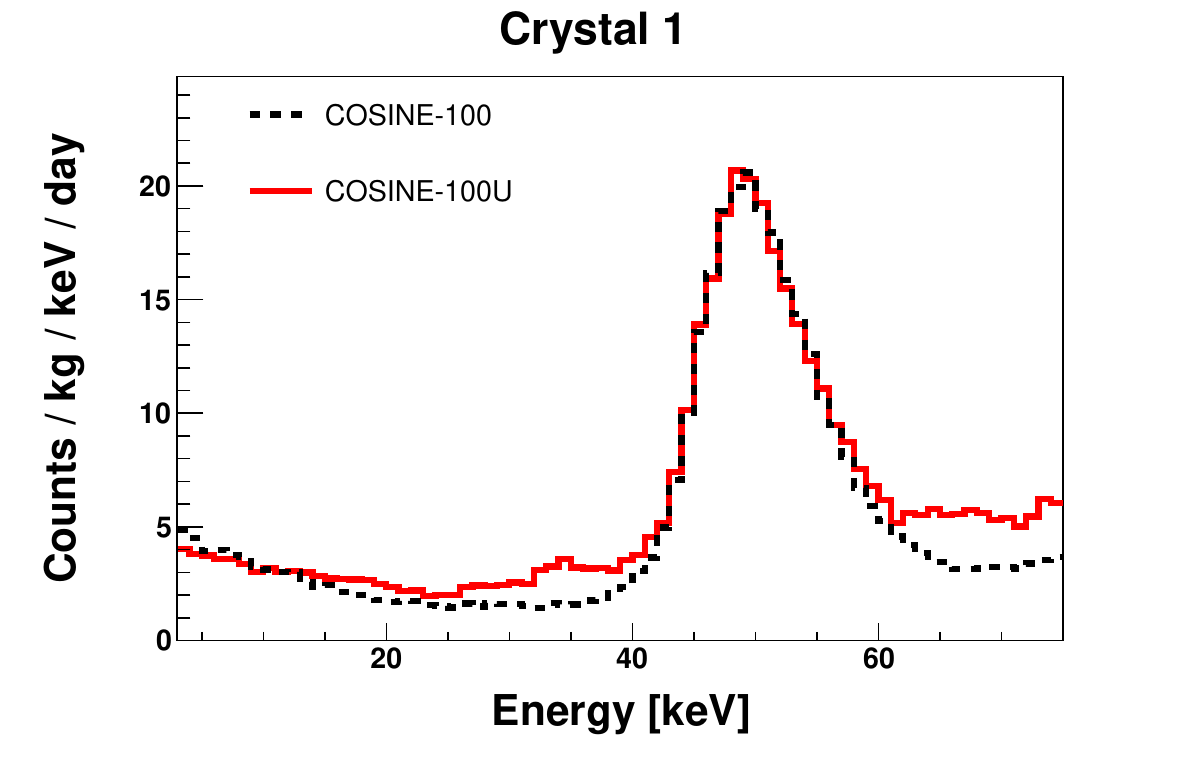}
   \includegraphics[width=0.23\textwidth]{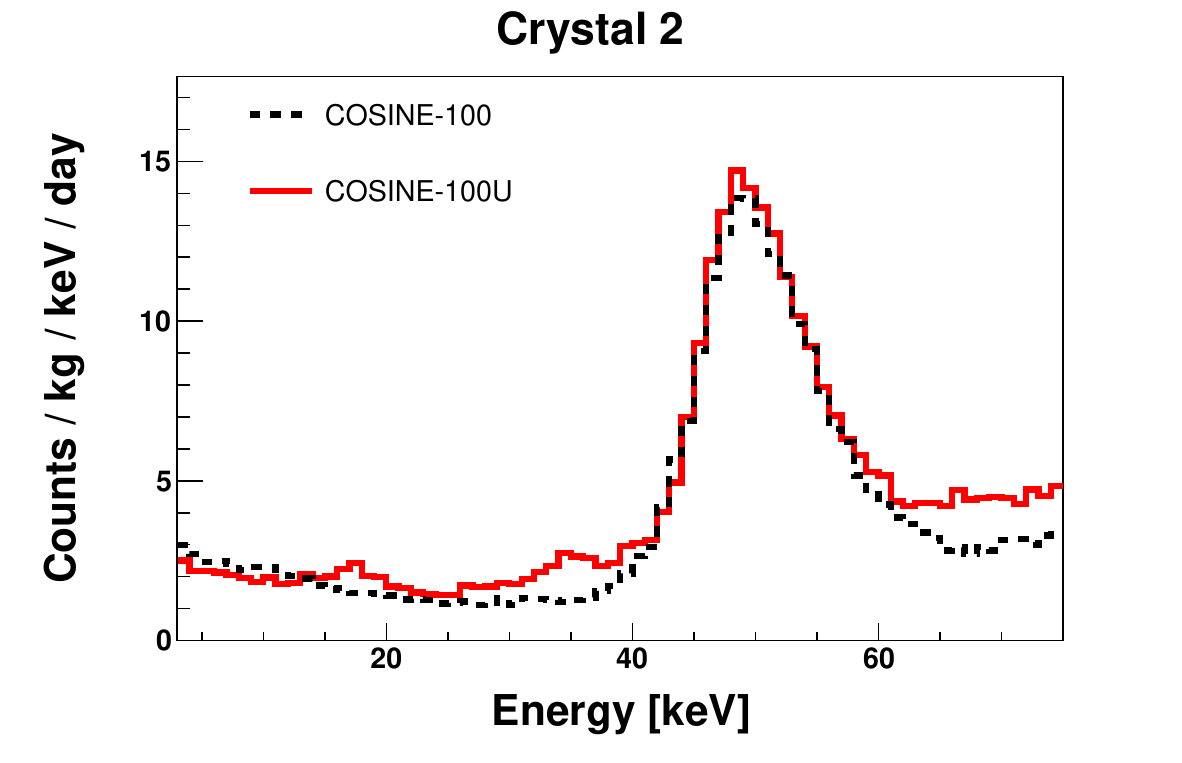}
   \includegraphics[width=0.23\textwidth]{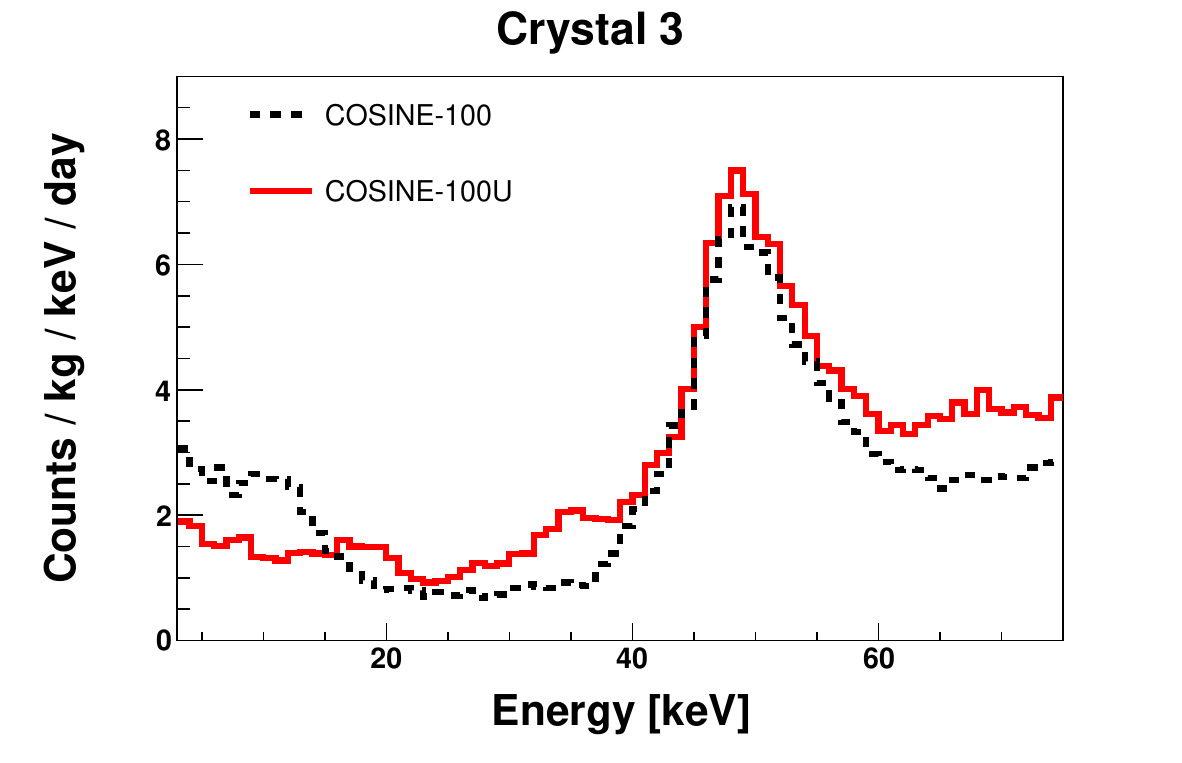}
   \includegraphics[width=0.23\textwidth]{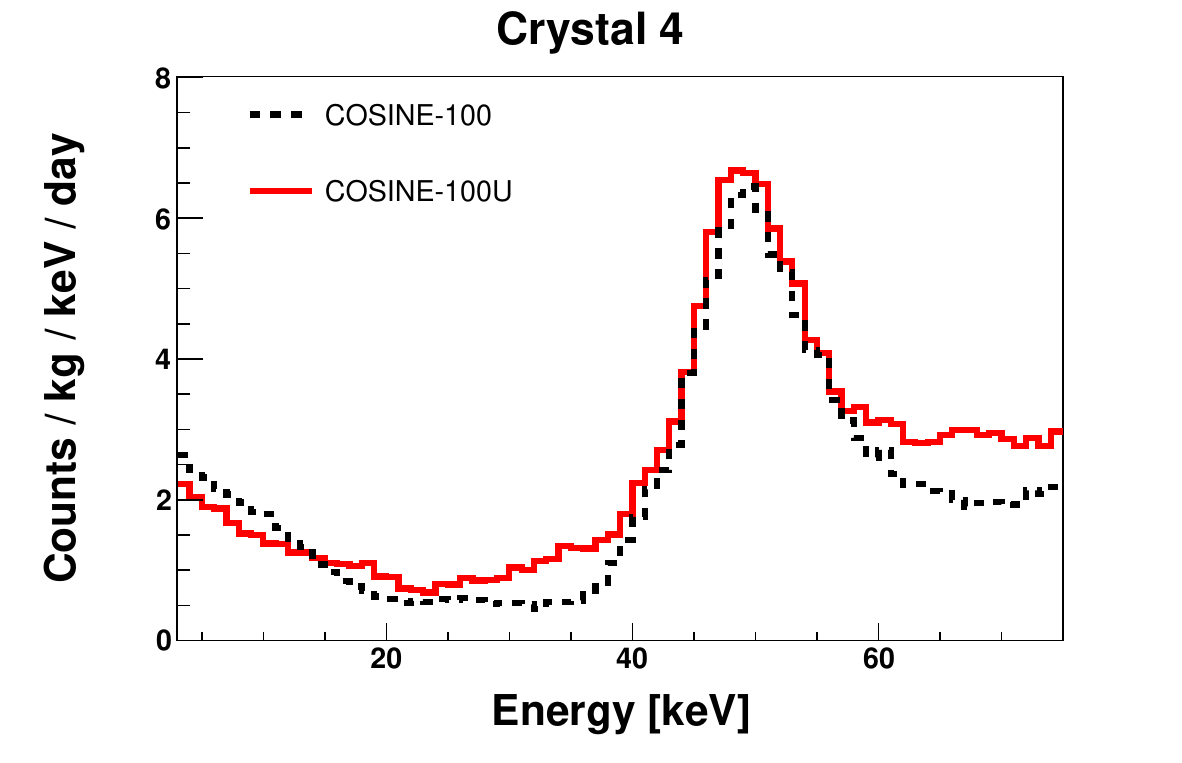}\\
   \includegraphics[width=0.23\textwidth]{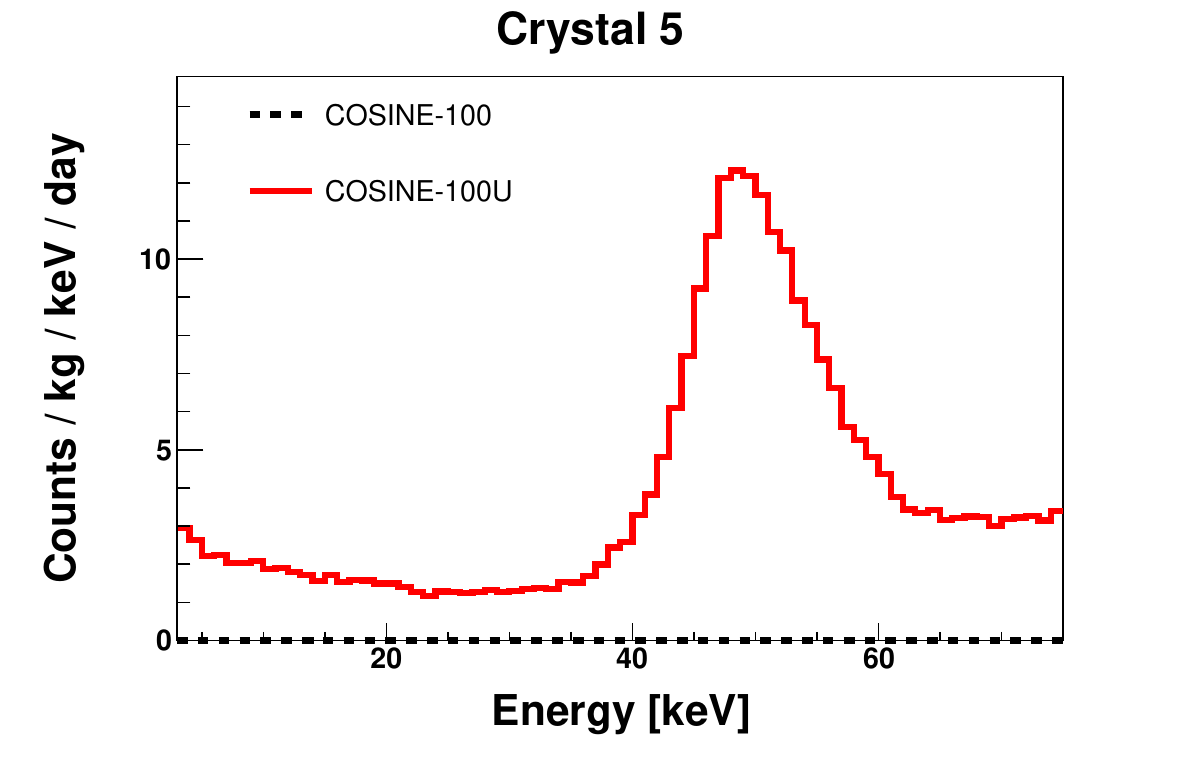}
   \includegraphics[width=0.23\textwidth]{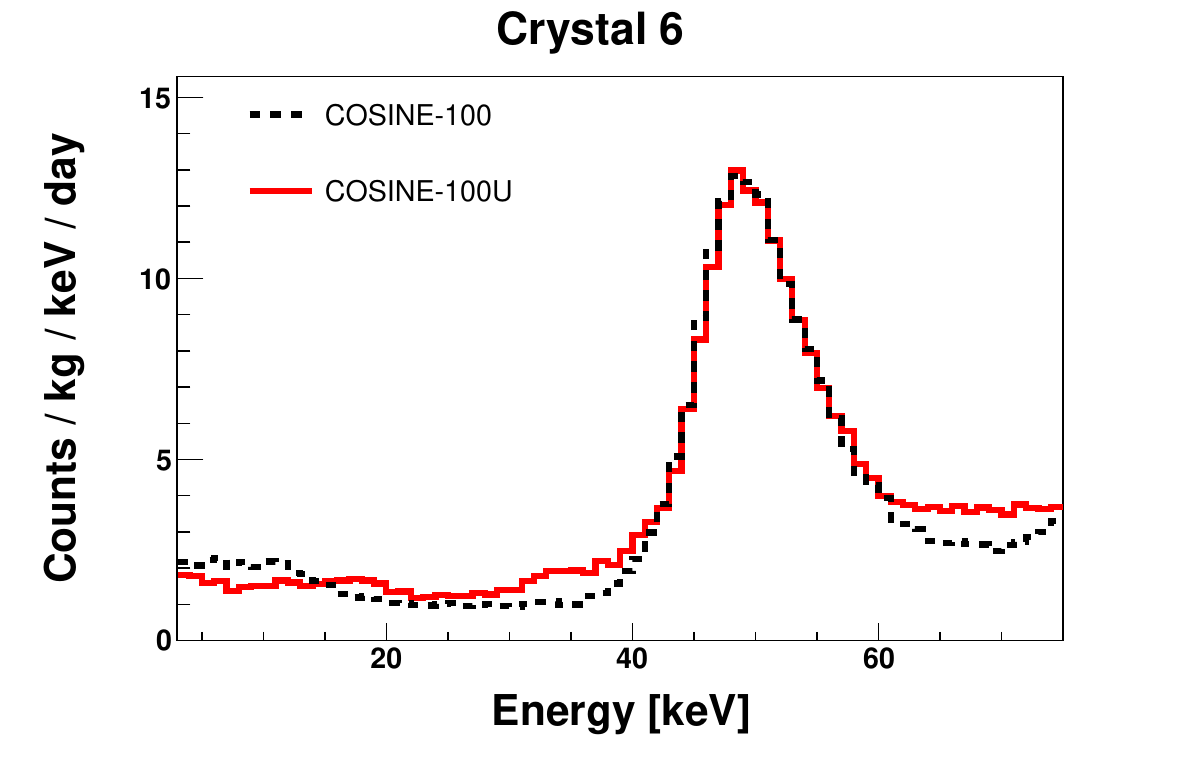}
   \includegraphics[width=0.23\textwidth]{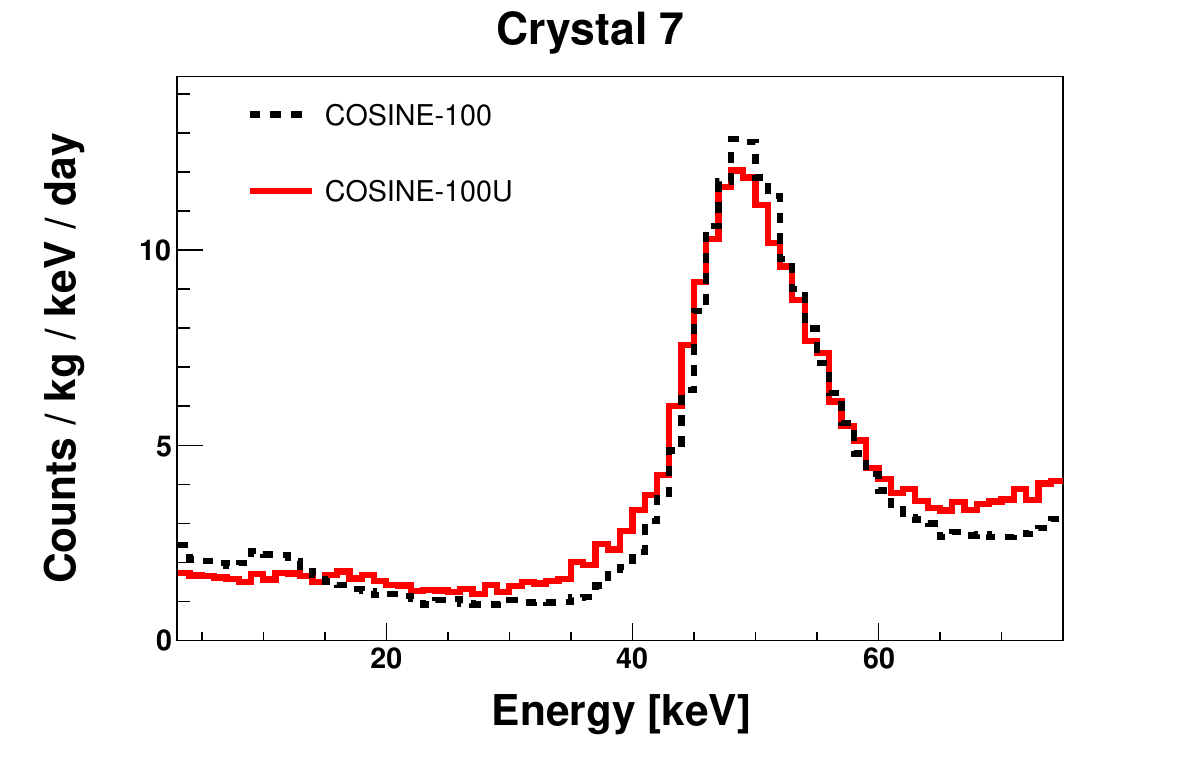}
   \includegraphics[width=0.23\textwidth]{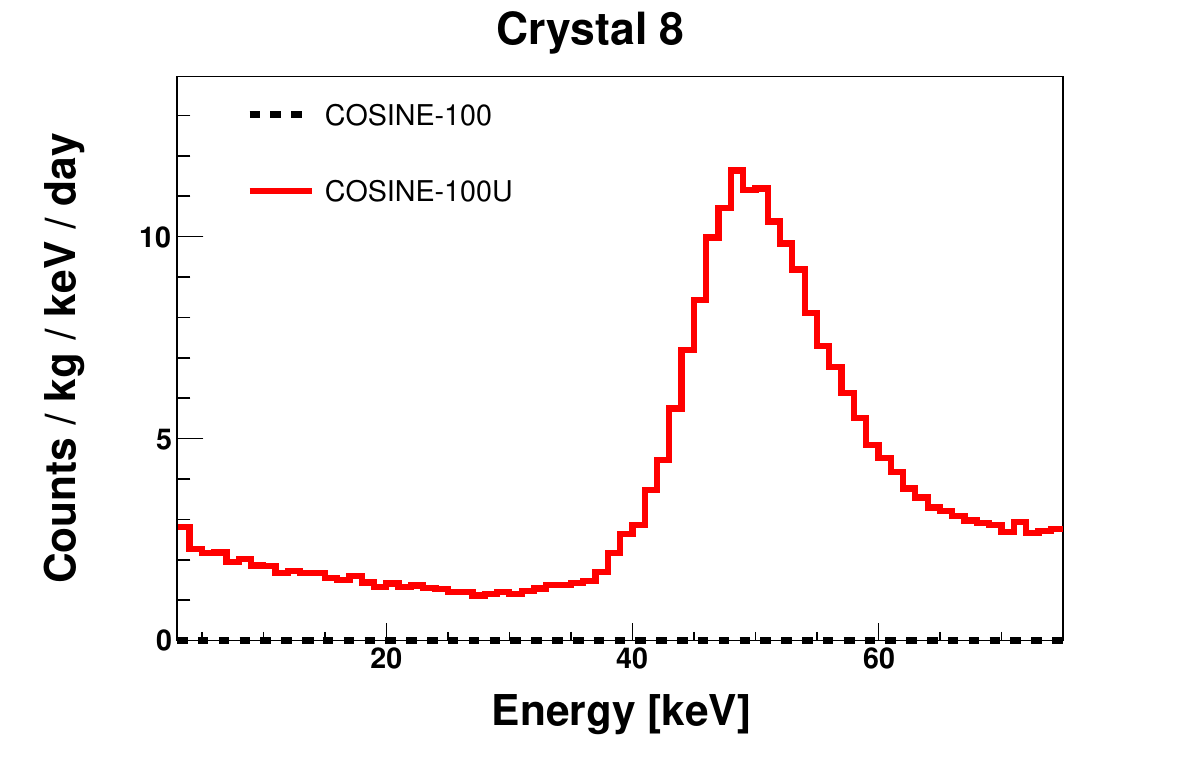}
   \caption{
\textbf{Comparison of the single-hit energy spectra for the eight NaI(Tl) crystals.}
The red solid curves show 2462~h (102.6~days) of room-temperature COSINE-100U data, and the black dashed curves show 698~h (29.1~days) of COSINE-100 data acquired near shutdown in March 2023. The same 80-keV LS veto threshold was applied to both data sets. Because Crystals~5 and 8 were excluded from the COSINE-100 physics analysis, no corresponding COSINE-100 spectra are shown for these detectors. For the remaining six crystals, the COSINE-100U spectra exhibit lower event rates below 15~keV, qualitatively consistent with reduced surface-related background contributions after re-encapsulation.
}
    \label{fig:singlehit}
\end{figure*}
Figure~\ref{fig:singlehit} compares the resulting single-hit
spectra after applying the same 80-keV LS-veto threshold to both
data sets and normalizing for their different live times. For the
six crystals with corresponding COSINE-100 spectra, COSINE-100U
shows lower event rates below 15~keV. This reduction is
qualitatively consistent with the lower surface-$\alpha$ activity
measured after re-encapsulation and with the known contribution of
surface-related components to the low-energy background~\cite{
COSINE-100:2024ola}. The increased light yield provides an
independent improvement by enabling a lower analysis threshold
and more effective discrimination of PMT-induced noise.

The sensitivity projection reported in
~\cite{lee2025upgrading} was based primarily on the detector
performance measured in testbench studies before installation in
Yemilab. The present measurements demonstrate that the fully
assembled COSINE-100U detector achieves high light yields,
reduced surface activity $\alpha$, and lower low-energy event
rates under actual underground operating conditions.
Although an updated dark-matter sensitivity calculation is beyond
the scope of this work, the in-situ detector performance reported
here indicates that the achievable sensitivity of COSINE-100U is
expected to surpass the earlier projection.

\section{Conclusion}
\label{sec:conclusion}

We developed and deployed an upgraded NaI(Tl) crystal encapsulation for the COSINE-100U experiment at Yemilab~\cite{Park:2024sio,lee2025upgrading}. The design removes the quartz optical windows used in COSINE-100 and couples the PMTs directly to beveled crystal end faces through thin silicone optical pads. The crystal--PMT assemblies are supported by PTFE inner structures and sealed inside OFC housings, providing both improved photon collection and protection from the surrounding liquid scintillator.

The room-temperature COSINE-100U data set demonstrates a substantial improvement in optical performance. All eight crystals show higher light yields than in COSINE-100, with values ranging from 15.8 to 27.7~p.e./keV; six crystals exceed 20~p.e./keV. The upgrade also restored Crystals~5 and 8, which had previously been excluded from the physics analysis because of degraded light collection. Compared with COSINE-100 data, the bulk-$\alpha$ rates are lower and consistent with the expected time evolution of internal $^{210}$Pb, while the surface-$\alpha$ rates are substantially reduced for all crystals with direct comparisons.

The combined improvements in light yield and low-energy background are expected to enhance the sensitivity of COSINE-100U to low-mass dark matter. Physics data taking at the nominal operating temperature of $-30\,^{\circ}\mathrm{C}$ began in May 2026. The low-temperature data will be used to quantify any additional gain in light yield, pulse-shape discrimination, and analysis threshold. These results establish a strong experimental basis for a more sensitive NaI(Tl)-based test of the DAMA/LIBRA annual-modulation claim.
\section*{Acknowledgment}
We thank the Korea Hydro and Nuclear Power (KHNP) Company for providing underground laboratory space at Yangyang and the IBS Research Solution Center (RSC) for providing high performance computing resources. 
This work is supported by:  the Institute for Basic Science (IBS) under project code IBS-R016-A1,  NRF-2021R1A2C3010989, NRF-2021R1A2C1013761, RS-2024-00356960, RS-2025-25442707 and RS-2025-16064659, Republic of Korea;
NSF Grants No. PHY-1913742, United States; 
STFC Grant ST/N000277/1 and ST/K001337/1, United Kingdom;
Grant No. 2021/06743-1, 2022/12002-7, 2022/13293-5 and 2025/01639-2 FAPESP, CAPES Finance Code 001, CNPq 304658/2023-5, Brazil;
UM Internal Grant non-APBN 2025, Indonesia.
\newpage

\bibliographystyle{IEEEtran}
\bibliography{dm}

\end{document}